# Digitalization of Free-Radical Polymerization

Elena F. Sheka

Institute of Physical Researches and Technology, Peoples' Friendship University of Russia (RUDN University), 117198 Moscow, Russia;

sheka@icp.ac.ru

**Abstract:** Current paper presents the first attempt of the digitalization of a complex polymerization process representing free-radical polymerization of vinyl monomers in the presence of small additives of stable radicals *TEMPO* and $C_{60}$ fullerene. Chain-reaction essence of the chemical process; quantum-chemical concept of elementary reactions; digital-twin presentation of their participants; virtual thermodynamic and kinetic descriptors of the latters; digital polymeryzation passports of individually composed reaction solutions lay the foundation of the digitalization carried out. Polymerization-passport predictions made for a number of solutions with different vinyl monomers and/or different initiating free radicals are fully consistent with chemical reality.

**Keywords:** digitalization of polymerization; free-radical polymerization; digital twins approach; energy graphs; thermodynamic and kinetic descriptors; polymerization passports; vinyl monomers; stable radicals; fullerene $C_{60}$

## 1. Introduction

Nowadays digitalization is increasingly being introduced into all spheres of human life and activity. Leaving aside the social and philosophical aspects of analyzing the results of this implementation, we cannot deny that, in general, society welcomes the latter. And the reason for this is the obvious progress in those areas in which this implementation is most significant. Science is still one of those types of human activity where the introduction of digitalization is just beginning. This does not mean the science of processing and using big data, but science in its original form - physics, chemistry, biology, etc., where digitalization concerns not only big data, but also the concepts that connect these data and allow you to work with them. Digitalization is taking its first steps in this direction. According to the author knowledge, the most advanced concepts in this area concern digitalization ones based on the general ideas of Digital Twins (DTs) [1,2] and Machine Learning (ML) algorithms [3]. However, despite the comparative transparency of both concepts, their implementation encounters many difficulties, the resolution of which requires turning to the maximum possible body of knowledge related to the subject of research. Obviously, processing such arrays requires a lot of time and labor, so the introduction of digitalization into scientific processes has been so far of a low speed, which, naturally, will quickly increases over time. To date, only the first results on this path are known [4-11], the main part of which relates to polymers. This paper provides the first summarized view on the digitalization of polymerization. The approach is based on a number of fundamental concepts among which there are the following. The pilot concept concerns the presentation of polymerization process as a chain reaction [12,13] involving a set of elementary reactions. The latters are considered as independent and superpositional, thereby allowing to use all the accumulated experience in the quantum chemical consideration of reactions [14-16]. Both initial reagents and final products of the reactions form the pool of digital twins that are the main objects of the virtual consideration

[8-11]. To turn the latter into digitalization instruments, two types of descriptors are introduced, thermodynamic and kinetic ones, which are equally characteristic of all elementary reactions. The unification of descriptors makes it possible to issue polymerization passports, which are a personal identifying document for each virtual reaction solution and provide a potential comparison of all the virtual characteristic features of the polymerization event under study with empirical reality. The efficiency of the approach is demonstrated in this article using the example of free-radical polymerization of vinyl monomers in the presence of stable radicals.

The paper is composed as following. Section 2 presents input data that lay the foundation of the digitalization of the chemical process under study. The data concern the grounds of Digital Twins concept, introduction and characterization of virtual reaction solutions, descriptors of the DTs polymerization and virtual device, and issuing of polymerization passports. The information relating to the initial stage of polymerization of vinyl monomers and its comparison with experimental data is given in Sections 3-5 for styrene, methyl methacrylate and *N*- isopropyl acrylamide (NIPA), respectively. Concluding comments are given in Section 6.

## 2. Input data to digitalization

### 2.1. Digital Twins Concept

A general algorithm of the DT concept can be presented schematically as [5,6]

*Digital twins → Virtual device → IT product.*

Scheme 1

Here, DTs are molecular models under study, virtual device is a carrier of a selected software, IT product covers a large set of computational results related to the DTs under different actions in the light of the software explored. The quality of IT product highly depends on how broadly and deeply the designed DTs pool covers the knowledge concerning the object under consideration and how adequate is the virtual device to peculiarities of this object. The first requirement can be met by a large set of the relevant DTs. As for the virtual device, it should not contradict with the object nature and perform calculations that provide obtaining reliable ITs.

Digital twins approach is completely free from statistical and random errors that accompany real experiments, and are favored to any modification. Received at the disposal of a virtual device, the DTs lay the foundation of a free independent virtual experiment in terms of the process of interest. As for the final IT product, an essential part of which is a comparative analysis of large number of the same-conditions received data, it mainly determines trends in the behavior of the object of interest. The above trinity lays the foundation of three pillars on which a further digitalization will rest.

### 2.2. Virtual reaction solutions

A predominant majority of polymerization processes, including free-radical polymerization (FRP) of vinyl monomers we are interested in, occur in particular reaction solutions. It would be logically to introduce a virtual reaction solution (VRS) as a source of DTs thus determining the first pillar of the input data. Since the DT concept the best works when revealing trends, a quite large variety of VRSs should be envisaged. A big amount of empirical data related to the FRP of vinyl monomers [17,18] as well to their free-radical copolymerization (FRCP) with stable radicals, such as *TEMPO* and $C_{60}$ fullerene [19-37] greatly facilitates the VRS modeling. As occurred, VRSs in all the cases consist of a similar set of ingredients, involving solvent, monomer, initiator as a source of free radicals, and small additives of stable ones. Evidently, each of the ingredients can largely vary, additionally strengthening the variation with different concentration, so possible variation of VRSs is large. This circumstance significantly complicates the selection of a limited number of compositions sufficient for the reliability of the obtained trend results, on the one hand, and the verification of the obtained trends when compared with experimental data, on the other. In light of this problem, when implementing the digitalization of polymerization [8-11], attention was drawn to a group of experiments on the kinetics of the initial stage of the FRP of vinyl monomer and their FRCP with stable radicals [36,37]. In this study, special attention is paid to the kinetics of the initial stage of the reaction, represented in terms of the time-dependent percentage monomer conversion $x(t)$. Experiments were performed in the same solvent, at the same temperature, in the same reactor, thus practically under the same conditions, including simultaneously changing monomers, initiators and stable radicals, while also keeping the concentrations of these ingredients fixed. Based on these experiments, a set of VRSs was proposed [8-11] in which the monomer is one of three, namely, of styrene, methyl methacrylate, and NIPA; the initiator is one of two – alkyl-nitrile *AIBN* or benzoyl peroxide (*BP*), and the stable radical is either *TEMPO* or $C_{60}$ fullerene, or both.

The next step of the VRSs concerns the concept of the polymerization process occurred in the VRS. The configuration of the playing field in this work is based on the chain-reaction concept laying the foundation of the complex polymerization process that thus presenting it as a well-traced sequence of superpositional elementary reactions [12, 13]. From the theoretical viewpoint, such a vision of polymerization is the most favorable for using the quantum-chemical (QC) techniques for its virtual consideration, reducing it to the consideration of individual elementary reactions [14-16]. The theory of elementary reactions and their QC consideration has been going on for many decades [38-41], and the only complaint about the certain limitations of these studies can be the fact that highly welcome QC calculations of sets of one-type reactions did not become widespread. The main novelty of the DT concept concerns just this key point since a large massive of elementary reactions, which is the playing field in the current study, allows clearly distinguish one-type reactions, performed under the same conditions, followed by a comparative analysis of their results, accompanied by the establishment of reliable trends. The status of 'the same conditions' implies the application to the same QC consideration, absolute temperature zero, and vacuum medium.

A rather complete list of the relevant elementary reactions related to the studied events is presented in Table 1. Nominations, listed in the table, concern simultaneously both reactions and their final products, As seen, the number of the corresponding objects is quite large so that the digitalization of the considered chemical process seems to be convincing enough. The first common characteristic of the reactions listed in the table is their radical character. However, they are therewith distinctly divided into two groups that cover association reactions (1) and (2), uniting free radicals with monomers, and grafting reactions (3)-(12) that in the case of the stable-radical $C_{60}$ are reactions of the fullerene derivatization of different kinds. Products of the first-group reactions as well as those of reactions (3b) and (4) are free radicals while those of the second group are either stable species or stable radicals in the case when $F$ presents either *TEMPO* or $C_{60}$, respectively. Over thirty years ago, the latter were called fullerenyls [42]. The past decades since then, this name has taken root [43-46] and we will use it in the future.

Reactions (1) and (2), uniting free radical with monomer $RM^\bullet$ and its oligomers $RM_n^\bullet$, evidently govern the FRP of monomer. The former is the cornerstone of the entire polymerization process, determining its feasibility as such. By selecting the most successful participants in this reaction empirically, the researchers opted for one-target free radicals such as $AIBN^\bullet$ and $BP^\bullet$, while the stable radical *TEMPO* was found unsuitable for this role. As for multi-target $C_{60}$, the first wave of polymer researchers, who introduced fullerene into polymerization and were confident in its radical nature, made repeated attempts to detect reaction (3b), accompanied by the growth of a polymer chain attached to the fullerene (reaction (4)) [19-32, 47]. The fate of reaction (3b) depends on the detailed configuration of the intermolecular junction between C$_{60}$ and a monomer. The latter is configured with two $sp^2$C-C bonds, one belonging to fullerene and the other presenting a vinyl group of monomer. Accordingly, the junction can be either two-dentate or one-dentate. If the first configuration causes the formation of a [2x2] cycloadded monoderivative stable radical $FM$ similar to the patterned C$_{60}$, the second results in the formation of fullerene-grafted monomer radical $FM^\bullet$ similar to $RM^\bullet$. Accordingly, reaction (2) describes the polymer chain growth $RM_n^\bullet$ initiated with a free radical while reaction (4) describes the monomer polymerization $FM_n^\bullet$, once grafted on fullerene. Although the existence of a reaction (3b) was suspected in a number of cases, a confident conclusion was not made and this reaction as well as reaction (4) were classified as unlikely. Nevertheless, we will take these reactions in what follows into account.

**Table 1**. Nomination of elementary reactions and/or digital twins related to the initial stage of the free-radical copolymerization of vinyl monomers with stable radicals

| Reaction mark | Reaction equation [1] | Reaction rate constant | Reaction type |
|---|---|---|---|
| (1) | $R^\bullet + M \to RM^\bullet$ | $k_i$ | generation of monomer-radicals |
| (2) | $RM^\bullet + (n-1)M \to RM_n^\bullet$ | $k_p$ | generation of oligomer-radicals, polymer chain growth |
| (3a) | $F + M \to FM$ | $k_{2m}^F$ | two-dentant grafting of monomer on C$_{60}$ |
| (3b) | $F + M \to FM^\bullet$ | $k_{1m}^F$ | one-dentant stable radical grafting of monomer, generation of monomer-radical |
| (4) | $FM^\bullet + (n-1)M \to FM_n^\bullet$ | $k_p^F$ | generation of oligomer-radical anchored to C$_{60}$, polymer chain growth |
| (5) | $S + M \to SM^\bullet \equiv SM$ | $k_{1m}^S$ | one-dentant coupling with monomer |
| (6) | $F + RM^\bullet \to FRM$ | $k_{rm}^F$ | monomer-radical grafting on C$_{60}$ |
| (7) | $S + RM^\bullet \to SRM$ | $k_{rm}^S$ | monomer-radical capturing with stable radical |
| (8) | $F + R^\bullet \to FR$ | $k_R^F$ | free radical grafting on C$_{60}$ |
| (9) | $S + R^\bullet \to SR$ | $k_R^F$ | free radical capturing with stable radical |
| (10) | $F + S \to FS$ | $k_S^F$ | stable radical grafting on C$_{60}$ |
| (11) | $R^\bullet + FM^\bullet \to RFM$ | $k_{FM}^R$ | monomer-radical $FM^\bullet$ capturing with free radical |
| (12) | $S + FM^\bullet \to SFM$ | $k_{FM}^S$ | monomer-radical $FM^\bullet$ capturing with radical S |

[1] $M, R, F, S$ mark a vinyl monomer, initiating free radicals (either $AIBN^\bullet$ or $BP^\bullet$), stable radicals (fullerene C$_{60}$, and TEMPO), respectively. Superscript black spot distinguishes radical participants of the relevant reactions.

Reaction (5) $SM$ reveals a capturing of monomer with one-target stable one. In contrast to the above fullerenyls, $SM$, when is formed, presents a routine non-radical one-bond-coupled intermolecular complex. In contrast to non-reactive monomer, the capturing of its monomer-radical $RM^\bullet$, described by reactions (6) $FRM$ and (7) $SRM$, is traditionally highly expected for both stable radicals. Actually, these reactions are of particular importance having the opportunity to completely stop the polymerization process. Then follow reactions (8) $FR$ and (9) $SR$, revealing a similar capturing of free radicals $R^\bullet$. Both reactions evidently affect the monomer polymerization, decreasing the number of initiating free radicals. Reaction (10) $FR$ takes into account the interaction of stable radicals between themselves, while reactions (11) $RFM$ and (12) $SFM$ describe the capturing of monomer-radical $FM^\bullet$ with stable ones. This set of elementary reactions is quite complete for the consideration of the initial stage of both FRP of vinyl monomers and their FRCP with stable radicals. The relevant DTs of their final products alongside with input ingredients form a large DTs pool, which is quite enough for the digitalization of the corresponding chemical process.

## 2.3. DT descriptors for polymerization and virtual device

The main goal of the digitalization of polymerization is to evaluate properties of the chemical process avoiding empirical synthesis and reducing expensive and time-demanding laboratory testing. To realize the wish it is necessary to build in silico models establishing a mathematical relationship between the structures of molecules and the considered properties. Molecular descriptors play a fundamental role in the action since they formally are the numerical representation of a molecular structure [48]. The DT concept is the most suitable for the job when aiming the second member of its trinity presented on Scheme 1 at performing the required molecular descriptors calculation. Evidently, a number of descriptors, depending on which namely face of the latter is considered, can characterize chemical process. In the current case, the matter is about FRP of vinyl monomers and their FRCP with additional stable radicals, concerning their initial stage, in particular, because of which our goal is to distinguish a proper set of molecular descriptors enabling to present the issue in the best way.

In the chain-reaction concept, each polymerization process is the result of a severe competition between bimolecular elementary reactions following their rates. Burdened with a large number of such reactions not only in the initial stage of the process, presented in Table 1, but many other throughout their entire propagation, any reaction solution resembles a 'pineapple- on- plantation', the individual scales of which represent windows of opportunity for individual reactions (see Figure 1a). Opening of these windows is controlled by the kinetics of the latters, with the fastest one being the winner. Thus, the reaction rate constants become the main virtual descriptors of the polymerization.

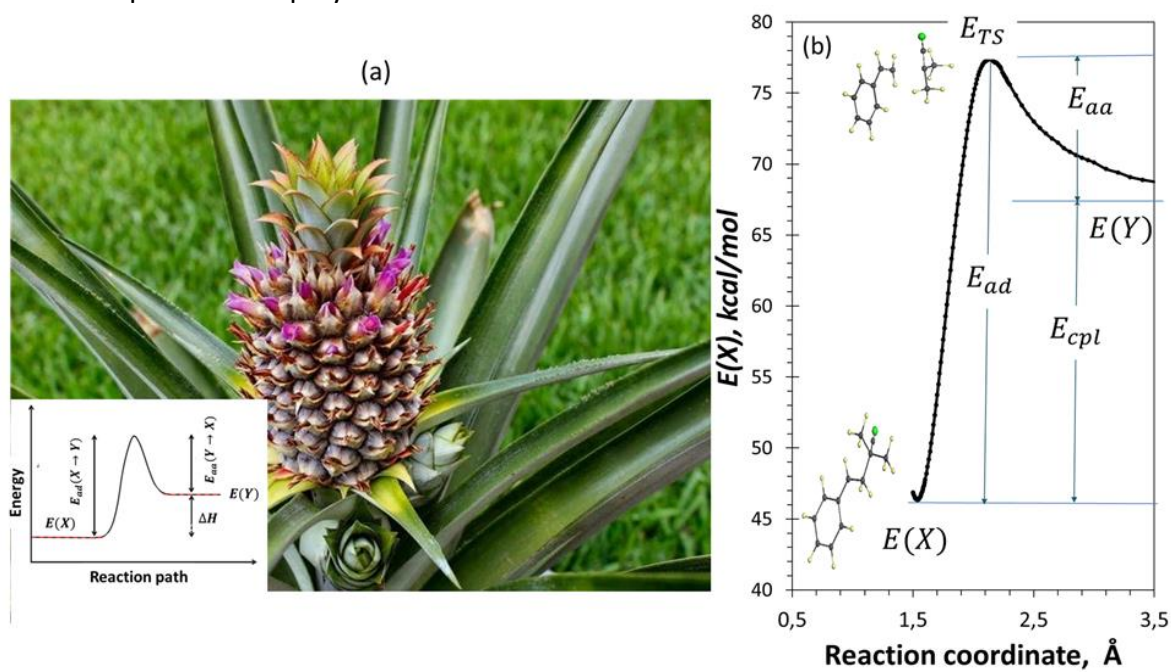

**Figure 1.** Universal energy graph of the pairwise intermolecular interaction in the system of many opportunities. (a) A 'pineapple-on-plantation' at polymerization. The photo from a free-accessible collection *pro-dachnikov.com* is used for the figure design. (b) Barrier profile of the decomposition of the monomer-radicals $RM^\bullet$ composed of styrene and free radical $AIBN^\bullet$. UHF AM1 calculations.

The elementary-reaction concept of FRP/FRCP provides one more important thing. This concerns a standard energy graph, applicable to this kind of reactions. Actually, each of them proceeds between a pair of reagents, the energy of intermolecular interaction between which follows a typical graph when the latters get closer in the chosen direction, taken as the reaction path or reaction coordinate (see Figure 1a). As seen in the figure, the graph includes the total energy of the reactants of the $Y$ pair, $E(Y)$, the energy of the final product of the pair constituents interaction, $E(X)$, and the energy of the transition state of the molecular complex under consideration, $E_{TS}(X \leftrightarrow Y)$ that determines the height of the reaction energetic barrier. There is also one more important energetic parameter – reaction enthalpy, $\Delta H$, or coupling energy, $E_{cpl}=E(X)-E(Y)$. Both quantities are directly related to the atomic structure of the object and in a number of cases, concerning the energetics of a chemical process, have every right to be considered as molecular descriptors. Thus, $E_{cpl}$ will play the role of a thermodynamic descriptor in what follows.

As for reaction kinetics, its standard description concerns the rate constant, $k(T)$, which is expressed through the Arrhenius relation [38-41]

$$k(T) = Ae^{\left(\frac{-E_a}{kT}\right)}. \qquad (1)$$

Here $A$ is a complex frequency factor, while $E_a$ presents the activation energy, which is either the energy of the $X$ product decomposition, $E_{ad}$, or the $Y$-pair constituents combination, $E_{ac}$. The main difficulty in evaluation of the $k(T)$ value is provided with highly complicated nature of frequency factor $A$. Its determination concerns basic problems of rotational-vibrational dynamics of polyatomic molecules, such as great number of both vibrational and rotational degrees of freedom as well as their anharmonicity. However, for one-type elementary reactions, $A$ is expected to change weakly [14, 38-41], so that activation energy becomes governing. Its value can be determined by building barrier profiles of either combination of the constituents of the molecular pairs $Y$ or decomposition of the final product $X$, respectively. The required barrier profiles can be calculated quantum chemically [49-51], one example of which is presented in Figure 1b. Calculations are more accurate when following the decomposition algorithm, starting with energy $E(X)$ at the equilibrium length of a covalent bond to be broken and following the bond regular elongation to its brake and above. As for the discussed elementary reactions, all of them proceed between the relevant DTs in the reverse direction because of which activation energies $E_{ac}$ are of the main interest. Accordingly, these values, renamed as $E_a$ to facilitate their comparison with available virtual and empirical data, will be used below as kinetic DT descriptors for FRP and/or FRCP of vinyl monomers.

Virtual device in the current study is represented with the CLUSTER-Z1 software [52,53] implementing AM1 version of the semi-empirical unrestricted Hartree-Fock (UHF) approach [54]. The program showed itself highly efficient concerning open-shell electronic systems such as fullerenes [55,56], graphene molecules [57], and stable radicals [58,59]. A detailed discussion concerning the choice of proper softwares for virtual FRP of vinyl monomers is presented elsewhere [8]. Digital twins of fullerenyls were designed basing on spin chemistry of fullerene $C_{60}$ [55,56].

**2.4. Digital polymerization passports**

A special matrix-tabulated format of the obtained results presentation concerns the IT product of the DT concept scheme and is the third pillar of the digitalization discussed. This format allows us to propose the issue of an universal "polymerization passport" (PP) of the VRS under study. The availability and high facility of computational methods makes it possible to issue such a passport to almost any VRS related to FRP and/or FRCP of vinyl monomers. The passports themselves, on the one hand, are strictly individual with respect to any of the VRS ingredients.

On the other hand, they represent an informative source of knowledge concerning the process under study. The first application of such document took place for the FRP of methyl methacrylate and its FRCP with fullerene C$_{60}$ and stable radical *TEMPO* [9]. Then detailed studies were performed for styrene [10] and *N-* isopropyl acrylamide (NIPA) [11]. Similarly to other identifying personal documents, virtual PPs consist of two pages. The first of them contains textual information, concerning the body, that involved the nomination of elementary reactions and the corresponding DTs supplemented with thermodynamic and kinetic DT descriptors, $E_{cpl}$ and $E_a$, respectively. The second page reveals photo-images of the considered DTs equilibrated structures. This PPs form occurred quite comfortable for the verification of the digitalized predictions with available empirical data.

## 3. Styrene polymerization

### 3.1. Digital polymerization passport of styrene

Table 2 and Figure 2 present two-page virtual PP of styrene. The passport is related to the FRP of styrene, initiated with free radical $AIBN^\bullet$ and transferred to the FRCP with small additives of *TEMPO* and C$_{60}$ fullerene. The table consists of three parts, listing the nominations, related to both elementary reactions and the relevant DTs that are important for the consideration of the FRP and FRCP under study; thermodynamic and kinetic DT descriptors, presented with $E_{cpl}$ and $E_a$ calculated values, respectively. All the digit data are obtained exploiting software CLUSTER-Z1, based on the semiempirical AM1 version of the unrestricted Hartree-Fock (UHF) approximation [52-54]. Bold $E_a$ values are evaluated when building barrier profiles of the relevant DTs decomposition [10] (a concise description of the barrier-profile methodology as well as the profiles themselves are given in the Supporting Information). 'Matrix elements' of the table are evidently divided into four groups marked with different colors. Yellow elements present elementary reactions alongside with the relevant DTs that govern FRP of styrene. Elements in light blue do the same but related to the FRCP of styrene with *TEMPO*, while faint-pink elements concern the FRCP of styrene with C$_{60}$. Elementary reaction (4) $FM_n^\bullet$ alongside with the related DTs, concerning the fullerene-stimulated polymerization of styrene, are highlighted in dark pink.

    Figure 2 represents the photo-page of the virtual PP and exhibits equilibrium structures of the DTs nominated in Table 2, thus visualizing a complexity of the considered VRS following the row arrangement of nominated DTs in the table. Thus, the table headings are presented in the figure by the image set (a). The set includes four main initial ingredients of the VRS, namely styrene ($M$), initiating free radical $AIBN^\bullet$ ($R^{A\bullet}$), stable radical *TEMPO* ($S$), and C$_{60}$ fullerene ($F$). The series is complemented with monomer-radical $R^A M^\bullet$ without which polymerization does not occur and the intermolecular interaction of which with initial ingredients determined how the polymerization proceeds. Radicals $R^A M^\bullet$, $R^{A\bullet}$ and *TEMPO* are one-target, reflecting the fact that the reactivity of each of them is provided with one electron located on a single atom. The atoms radical efficiency is determined by the number of unpaired electrons $N_{DA}$ [55,56] which plays the role of the atom reactivity descriptor. These atoms, carbons in the case of $R^A M^\bullet$ and $R^{A\bullet}$ as well as oxygen in $S^\bullet$, are highlighted in red. In contrast, radical efficiency of $F$ is distributed over 60 atoms of the C$_{60}$ molecule, remaining the same within each of five selected groups of atoms but varying between the groups, which is highlighted in different colors in the insert [60]. This multi-color image evidences the multi-target ability of the molecule, whose 12 light gray atoms are the most active, once characterized by the largest reactivity descriptor $N_{DA}$. All the fullerenyls discussed in the paper present fullerene monoderivatives, designed by attaching each corresponding addends to the same carbon atom from the light gray group.

Digital twins of the first row of Table 2 are exhibited in Figure 2 by image sets (b) and (c). The former represents a growth of the styrene polymer chain via a continuous length increasing of the styrene oligomer-radicals $R^A M_n^\bullet$, described in details earlier [8]. The next set unites the remaining four matrix elements of the row, two of which, namely $SM$ and $FM$, are standard stable intermolecular complexes, while $R^A M^\bullet$ and $FM^\bullet$ are radicals with the strongest descriptors of reactivity of $N_{DA}$ = 0.67 $e$ on carbon atoms of the styrene vinyl group in both cases.

**Table 2.** Elementary reactions and DTs of their final products supplemented with thermodynamic and kinetic descriptors related to the FRCP of styrene with stable radicals *TEMPO* and C$_{60}$, while initiated with $AIBN^\bullet$ free radical

|  | $M$ | $R^A M^\bullet$ | $AIBN^\bullet$ ($R^{A\bullet}$) | $TEMPO$ ($S^\bullet$) | $C_{60}$ ($F$) | |
|---|---|---|---|---|---|---|
|  |  |  |  |  | 2-dentate | 1-dentate |
| **Digital Twins' set** | | | | | | |
| $M$ | $M_n$ | $R^A M_n^{\bullet\bullet}$ | $R^A M^\bullet$ | $SM$ | $FM$ | $FM^\bullet$ |
| $R^A M^\bullet$ | - | $(R^A M)_2$ | $R^A R^A M$ | $SR^A M$ | $FR^A M$ | |
| $R^{A\bullet}$ | - | - | - | $SR^A$ | $FR^A$ | |
| $S^\bullet$ | - | - | - | - | $FS$ | |
| $FM^\bullet$ | $FM_n^{\bullet\bullet}$ | - | $R^A FM$ | $SFM$ | - | |
| **Thermodynamic descriptors $E_{cpl}$, kcal/mol** [1] | | | | | | |
| $M$ |  | -7.38 ÷ -29.01 (2-6) | −22.51 (1) | 4.58 | -34.31 | -18.59 (1) |
| $R^A M^\bullet$ | - |  | -29.80 | -5.03 | -19.11 | |
| $R^{A\bullet}$ | - | - | - | 5.01 | -20.14 | |
| $S^\bullet$ | - | - | - | - | 0.98 | |
| $FM^\bullet$ | -13.86 (2) -18.81 (3) -18.88 (4) -23.04 (5) | - | -19.63 | -0.89 | - | |
| **Kinetic descriptors $E_a$, kcal/mol** [1,2] | | | | | | |
| $M$ |  | 12.06 (2) 6.12 - 16.49 [3] (2-6) | 8.50 (1) | 19.28 | 24.52 | 8.38 (1) |
| $R^A M^\bullet$ | - | - | not defined | 19.04 | 9.73 | |
| $R^{A\bullet}$ | - | - | - | 18.74 | 9.41 | |
| $S^\bullet$ | - | - | - | - | not combined | |
| $FM^\bullet$ | 11.25 (2) | - | not defined | 19.54 | - | |

[1] Digits in brackets mark the number of monomers in the oligomer chain.
[2] Bold data are determined from the decomposition barrier profiles presented in Figures S1 and S2.
[3] The data are calculated by using Evans-Polany-Semenov relation presented in [8].

Four DTs of the table's second row form the image set (d) in the figure. The series presents all the cases of potential capturing of monomer-radical $R^A M^\bullet$, which leads to the termination of the polymer chain $R^A M_n^\bullet$ growth. Two first members of the series, considered in details earlier [8], can be attributed to self-inhibiting. However, the very fact of successful and almost instantaneous empirical FRP of styrene indicates the negligible role of these participants in the polymerization process, so they will not be considered further. Single DTs remaining in the third and fourth rows of the table form the image set (e). And, finally, the image set (f) $FM_n^\bullet$, located in the last row of the table, completes the photo-page of the styrene virtual PP.

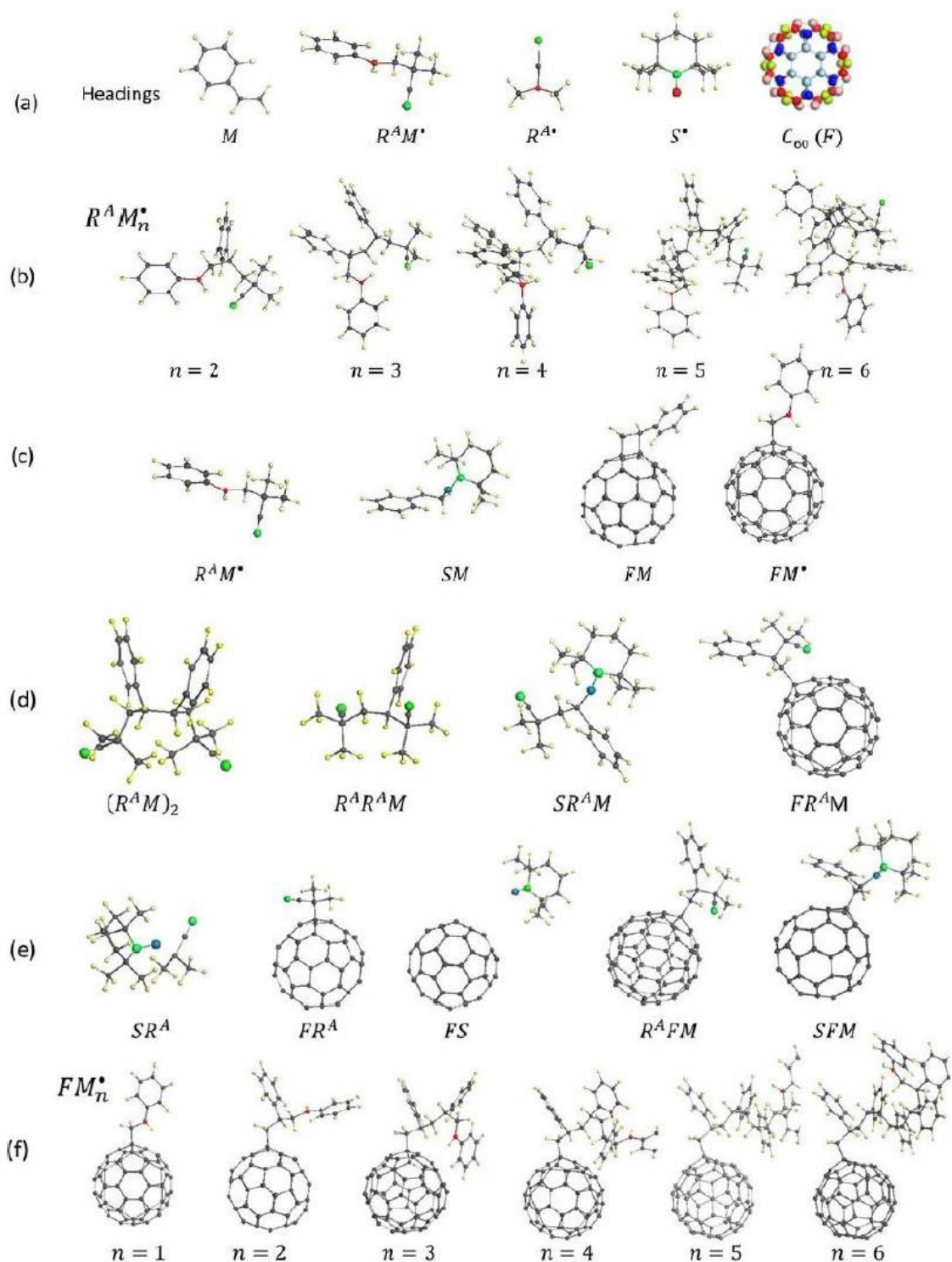

**Figure 2.** Equilibrium structures of digital twins related to the FRCP of styrene with stable radicals. (a) Structures, representing headings of Table 2. Colored image of $C_{60}$ exhibits a specific spin density distribution over the molecule carbon atoms [55,56]. (b) Oligomer radicals $R^A M_n^\bullet$ for $n$ from 2 till 6. (c) Digital twins produced by intermolecular interaction of styrene with free radical $R^{A\bullet}$, *TEMPO* and $C_{60}$, both two-dentant and one-dentant attached. (d) DTs describing the interaction of monomer-radical $R^A M^\bullet$ with itself, free radical $R^{A\bullet}$, *TEMPO*, and fullerene $C_{60}$, respectively; (e) DTs related to the interaction of radicals between themselves - $R^{A\bullet}$ with *TEMPO* and fullerene, fullerene with *TEMPO*, $R^{A\bullet}$ and *TEMPO* with $FM^\bullet$; (f) Oligomer radicals $F(M)_n^\bullet$ for $n$ from 1 till 6. Small yellow and gray balls mark hydrogen and carbon atoms, respectively. Larger green and blue balls depict nitrogen and oxygen atoms. Red balls mark carbon (small) and oxygen (large) target atoms. UHF AM1 calculations

**Table 3.** Elementary reactions and DTs of their final products supplemented with thermodynamic and kinetic descriptors related to the FRCP of styrene with stable radicals *TEMPO* and $C_{60}$, while initiated with $BP^\bullet$ free radical

| | $M$ | $R^P M^\bullet$ | $BP^\bullet$ $(R^{P\bullet})$ | $TEMPO$ $(S^\bullet)$ | $C_{60}$ $(F)$ | |
|---|---|---|---|---|---|---|
| | **Digital Twins' set** | | | | | |
| | | | | | *2-dentate* | *1-dentate* |
| $M$ | $M_n$ | $R^P M_n^{\bullet\bullet}$ | $R^P M^\bullet$ | $SM^\bullet$ | $FM$ | $FM^\bullet$ |
| $R^P M^\bullet$ | - | $(R^P M)_2$ | $R^P R^P M$ | $SR^P M$ | $FR^P M$ | |
| $R^{P\bullet}$ | - | - | | $SR^P$ | $FR^P$ | |
| $S^\bullet$ | - | - | - | - | $FS$ | |
| $FM^\bullet$ | $FM_n^{\bullet\bullet}$ | - | $R^P FM$ | $SFM$ | | |
| | **Thermodynamic descriptors $E_{cpl}$, kcal/mol** [1] | | | | | |
| $M$ | | -23.23 (2) / -18.47 (3) | -40.33 (1) | 4.58 | -34.31 | -18.59 (1) |
| $R^P M^\bullet$ | | | | -2.85 | --23.12 | |
| $R^{P\bullet}$ | | | | --0.63 | -40.62 | |
| $S^\bullet$ | | | | - | 0.98 | |
| $FM^\bullet$ | -13.862 (2) / -18.813 (3) / -18,879 (4) / -23.042 (5) | - | -19.63 | -0.89 | | |
| | **Kinetic descriptors $E_a$, kcal/mol** [1,2] | | | | | |
| $M$ | | **8.78 (2)** | **2.79 (1)** | 19.28 | 24.52 | **8.38 (1)** |
| $R^P M^\bullet$ | | | | 17.60 | 7.02 | |
| $R^{P\bullet}$ | | | | not coupled | 28.82 | |
| $S^\bullet$ | | | | - | not grafted | |
| $FM^\bullet$ | **11.25 (2)** | - | not determined | 19.54 | | |

[1] Digits in brackets mark the number of monomers in the oligomer chain.
[2] Bold data are determined from the decomposition barrier profiles presented in Figures S3.

As can be seen from the above, visualization of DTs significantly enlivens inhabitants of the VRS and makes their analysis more presentive. At the same time, it becomes the best way to trace the polymerization process when replacing any participant in this process by another one as well as to suggest the first predictions. As for the role of free radicals, Table 3 and Figure 3 present the styrene virtual PP when alkyl-nitrile $AIBN^\bullet$ is substituted with benzoyl-peroxide $BP^\bullet(R^{P\bullet})$. Not repeating the photo-page of the previous virtual PP, Figure 3 presents only those DTs that react on the free radical replacement. As seen in the figure, the first innovation concerns the monomer-radical $R^P M^\bullet$ itself and the resulting polymer chain $R^P M_n^\bullet$. This virtual chain,

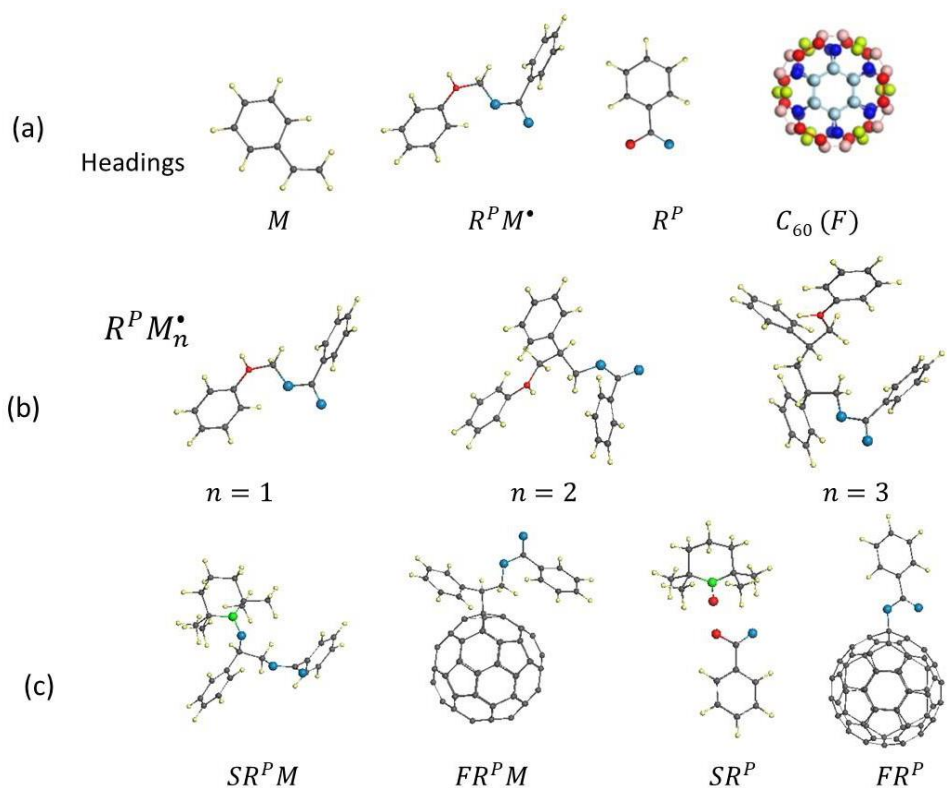

**Figure 3.** Equilibrium structures of digital twins of FRCP of styrene with stable radicals. (a) Structures, representing headings of Table 3. (b) Oligomer radicals $R^P M^{\bullet}_{n+1}$ for $n$ from 1 till 3. (c) Digital twins produced by intermolecular interaction of styrene monomer-radical $R^P M^{\bullet}$ with TEMPO and $C_{60}$, as well as free radical $R^P$ with TEMPO and fullerene. Small yellow and gray balls mark hydrogen and carbon atoms, respectively. Larger green and blue balls depict nitrogen and oxygen atoms. Red balls mark carbon (small) and oxygen (large) target atoms. UHF AM1 calculations

designed recently [10], is presented by the image set (b) in in the figure and consists of monomer-radical and two oligomer ones. Generally, it has much in common with that one previously initiated by the $AIBN^{\bullet}$ [8], while, naturally, the structural details of the chain, determined by intermolecular junctions as well as by their composition in space, are quite different. The image set (c) in the figure displays results of capturing of either monomer-radical $R^P M^{\bullet}$ or free radical $R^{P\bullet}$ with stable radicals TEMPO and $C_{60}$.

### 3.2. Virtual and real kinetics of the styrene polymerization

Figure 4 accumulates empirical data related to the time-dependent percentage conversion $x(t)$ of monomer that well represents kinetics of the initial stage of the FRP of styrene and its FRCP with fullerene $C_{60}$ and TEMPO, while being initiated with either $AIBN^{\bullet}$ or $BP^{\bullet}$ free radicals [36,37]. The panorama presents a complex of experiments performed under the same conditions concerning the temperature, solvent, contents of monomer, chemical content of free and stable radicals. Graphs 1 in all the panels present the referent FRP of styrene. Figure 4a exhibits the effect of small additives of $C_{60}$ on the referent process. Figure 4b shows effect of combined action of $C_{60}$ and TEMPO, Figure 4c does the same concerning the replacement of $AIBN^{\bullet}$ with $BP^{\bullet}$ followed with small additive of $C_{60}$. The observed effects are well pronounced and strong. Kinetic descriptors $E_a$ listed in Tables 2 and 3 do not pretend on the reproduction of the empirical monomer-conversion dependences discussed above, but they allow identifying key reasons for their behavior.

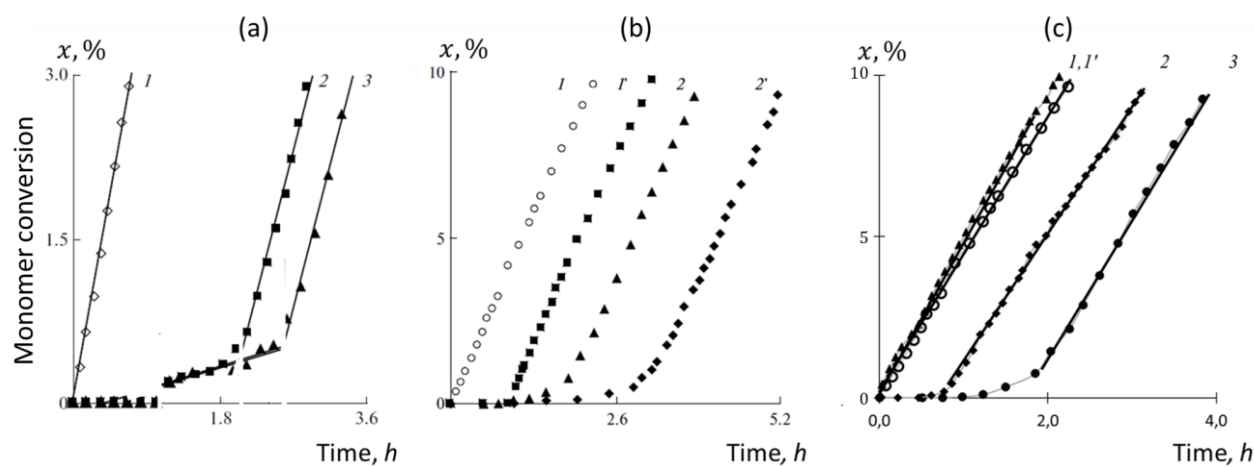

**Figure 4.** Empirical kinetics of the initial stage of both FRP of styrene and its FRCP with *TEMPO* and fullerene $C_{60}$. (a) $AIBN^{\bullet}$-initiated conversion of styrene in the presence of different [$C_{60}$]: 0 (graph 1); 1.0 × $10^{-3}$ (graph 2); 2.0 × $10^{-3}$ mol/L (graph 3). (b) The same but with [*TEMPO* ($C_{60}$)] 0 (graph 1); [*TEMPO*] 1.0 × $10^{-3}$ mol/L and [$C_{60}$] 0 (graph 1'); [*TEMPO*] 0 and [$C_{60}$] 1.0 × $10^{-3}$ mol/L (graph 2); [*TEMPO* ($C_{60}$)] 1.0 × $10^{-3}$ mol/L (graph 2'). (c) Conversion of styrene, initiated with $AIBN^{\bullet}$ (graphs 1' and 3) and $BP^{\bullet}$(graphs 1 and 2) in the absence (1' and 1) and in the presence (3 and 2) of fullerene [$C_{60}$]=2.0 mol/L. T=60⁰C; o-DCB solvent; [St]=2.0 mol/L; [AIBN (BP)]=2.0x$10^{-2}$mol/L. Digitalized data of Refs. 36 and 37.

    Let us start with Figure 4a. As stated by the chemists [36,37], observed empirical effect concerns mainly the generation of an induction period (IP) caused by the fullerene addition. Therewith, the IP duration elongates with increasing fullerene content, and the conversion rate of the monomer massive polymerization, occurred after the IP termination, remains practically identical to the referent one. Taken together these features were interpreted by the inhibition action of fullerene followed with its full consumption occurred by the IP end. Neither type of the action nor clearly observed slope of the IP graph were interpreted unambiguously.

    Applying to the styrene PP data, the figure content is related to the elementary reactions and the relevant DTs marked by yellow and faint-pink in Table 2. As seen from the table, the FRP is governed with $E_a$ descriptors filling the range from 6.1 to 16.5 kcal/mol. The values are typical for vinyl monomers [8] and *post factum* are kinetically quite favorable for their FRP to be successful. The corresponding reactions include the initiation of the monomer-radical $R^A M^{\bullet}$ and the successive growth of the polymer chain $R^A M_n^{\bullet}$. As for the effect caused by $C_{60}$, three reactions, involving the formation of the $FM^{\bullet}$ monomer-radical and two stable fullerenyl radicals $FR^A M$ and $FR^A$, should be considered. Generally, listed $E_a$ data are not enough to judge which constant rate is the fastest, so that frequency factors $A$ evidently are needed. However, in the case of similar reactions, which is the case of reactions (3b), (6), and (8) (see Table 1), $E_a$ descriptors can be confidently used for the interrelation between the latters to be suggested. The following reaction-constant series looks like

$$k_{1m}^F > k_R^F > k_{rm}^F, \qquad (2)$$

indicating that polymerization of styrene on fullerene is the fastest.

    The relation $k_{1m}^F > k_R^F$ is supported empirically when studying the $C_{60}$ conversion in the presence of either both styrene and $AIBN$ or $AIBN$ only, thus exhibiting a fivefold decreasing of the rate in the second case, which reveals a drastic difference in $FR^A$ and $FR^A M$ reactions.

Moreover, empirically observed $k_R^F$ rate in the $AIBN$ solution remains unchanged when not styrene, but methyl methacrylate is input [37] thus pointing to its independence from the type of monomers. As will be shown in Section 4.2, the $FR^A$ reaction dominates in the case of the FRCP of methyl methacrylate with C$_{60}$. Therewith, no IP but a significant decreasing in the referent conversion-graph slope is observed when fullerene is input in the reaction solution. Therefore, in contrast to methyl methacrylate, as follows from Figure 4a, $FR^A$ reaction does not govern the FRCP of styrene.

Thus, the reasons of the empirically observed IP in Figure 4a and, consequently, the fate of styrene polymerization on fullerene are decided with the relation between the rate constants $k_{1m}^F$ and $k_i$ for reactions $FM^\bullet$ and $R^AM^\bullet$, respectively. As follows from Table 2, $E_a$ descriptor for $FM^\bullet$ only slightly exceeds that for $R^AM^\bullet$,. However, pre-exponential frequency factor $A$ is reasonably expected to be bigger for fullerenyl than that of much less monomer-radical $R^AM^\bullet$, thus strengthening the required inequality $k_{1m}^F > k_i$. Accordingly, we have to conclude that the input of fullerene into VRS, intended for FRP of styrene, cancels $R^AM^\bullet$ reaction, replacing it with $FM^\bullet$, thus starting polymerization of the monomer driven by the monomer-radical $FM^\bullet$. This explains a drastic consumption of C$_{60}$ in the styrene-load solution during the initial stage of the polymerization [37], thus allows concluding that reaction $FM_n^\bullet$ governs this stage because of which for styrene monomers

$$k_{1m}^F > k_i$$
$$k_{1m}^F > k_R^F \ . \quad (3)$$

Naturally, the three orders of magnitude difference in the contents of styrene and C$_{60}$ practically suppresses the $FM_n^\bullet$ conversion dependence against the background of that for $R^AM_n^\bullet$ to almost zero, which makes its appearance in the overall picture of monomer conversion $x(t)$ in Figure 4a similar to close-to-zero IP. As seen, the monotonically increasing conversion initially occurs spasmodically. This behavior, however, is within experimental errors. Graph 2, repeated later in the next experiments, presented in Figures 4b and 4c (graphs 2 and 3, respectively), has a well-developed expected cross-sectional appearance of a rounded hockey stick. Therefore, the FRCP of styrene with C$_{60}$ starts with the polymerization of styrene, both stimulated with and anchored at fullerene. When all fullerene molecules are consumed, a FRP of styrene, initiated with a free radical $AIBN^\bullet$ proceeds and its conversion graph is identical to the referent one.

Let us move on to the situation when *TEMPO* is input to the reaction solution together with styrene and $AIBN$. As can be seen from Figure 4b (graphs 1 and 1'), the addition of *TEMPO* is accompanied by the appearance of a classical IP graph with almost zero slope to the coordinate axis. Upon reaching the end of this period, the FRP styrene begins, whose conversion is practically identical to the referent one. In Table 2, elementary reactions concerning *TEMPO* are presented by cells marked in light blue. As follows from the table, three reactions, namely, $SM$, $SR^AM$ and $SR^A$ are possible, markedly differing by thermodynamic $E_{cpl}$ descriptors. Two of them ($SM$ and $SR^A$) are characterized by positive $E_{cpl}$ pointing to the relation $E_{ad} > E_{ac}$. It is difficult to imagine that these reactions can be effective in a complex polymerization medium, full of competing players. In contrast to the above two, reaction $SR^AM$ is evidently thermodynamically possible. Under these conditions, a choice in favor of the $SR^AM$ reaction is made taking into account both $E_a$ and $E_{cpl}$ descriptors. This reaction corresponds to the capturing of monomer-radicals $R^AM^\bullet$ with *TEMPO*, terminates the FRP of styrene and confidently explains the presence of IP on the monomer-conversion curve with a zero slope relative to the abscissa.

Nevertheless, a question remains, why reaction $SR^AM$ with $E_a$= 19.04 kcal/mol precedes reactions $R^AM_n^\bullet$ with smaller kinetic descriptor, which does not allow the FRP of styrene to propagate when the monomer-radical is formed. Here we meet again the case when not only $E_a$ descriptor, but a pre-exponential factor $A$ in Eq. (1) should be taken into account. Nevertheless, it is possible to reliably conclude that, basing on both empirical and virtual realities, for styrene monomers

$$k_{rm}^S > k_p. \tag{4}$$

When suggesting that all the elementary reactions are superpositional and occur independently on each other, input of small addition of $C_{60}$ to the discussed reaction solution does not disturb both reactions themselves discussed above as well as interrelation between them, but just add some new reactive opportunities. Since the presence of *TEMPO* cannot influence the formation of monomer-radical $FM^\bullet$ and $R^A M^\bullet$ as well as the relationship between the rate constants of the corresponding reactions, the ratio of $k_{FM}^S$ and $k_p^F$ decides the fate of the beginning of the FRCP of styrene with *TEMPO* and $C_{60}$. Virtual data do not allow evaluating the requested ratio, while experiment evidences in favor of $SFM$ one. Actually, attentive analysis shows that graph 2' in Figure 4b consists of three parts. The first coincides with the IP period of the FRCP of styrene with *TEMPO* (graph 1') and evidences that the rate constant $k_{FM}^S$ is the biggest. The IP is ended when all the *TEMPO* molecules are consumed that is why its length coincides with that one related to graph 1'. When $SFM$ reaction is terminated, the reaction $FM_n^\bullet$ starts just reproducing graph 2. The corresponding conversion is continued with $R^A M_n^\bullet$ reaction, so that graph 2' copies graph 2, but shifted along the axe on the length of the IP provided with *TEMPO*, analogously to the joint behavior of graphs 1 and 1'. Accordingly, the rate-constant series for the FRCP of styrene with *TEMPO* and $C_{60}$, expressed in Eqs. (2)-(4), is completed with one more

$$k_{FM}^S > k_p^F. \tag{5}$$

As mentioned earlier, any of the conclusions, similar to the above made, are valid for a particular chemical contents of the reaction solution. Actually, it becomes evident when alkyl-nitrile $AIBN$ is replaced with benzoyl peroxide. Virtual data related to empirical results shown in Figure 4c are listed in Table 3. Analyzing these data and reasoning as above, it is easy to conclude the following. Kinetic descriptors $E_a$ of elementary reactions $R^P M^\bullet$ and $R^P M_n^\bullet$, although different for $R^A M^\bullet$ and $R^A M_n^\bullet$ from Table 2, are within acceptable values for the FRP of vinyl monomers [8]. This conclusion is fully confirmed by the close proximity of referent graphs 1 and 1' in the figure.

$E_a$ descriptors allow suggesting that fullerene-provided reactions form a set, rate-constant members of which follow the sequence

$$k_{rm}^F > k_{1m}^F \gg k_R^F. \tag{6}$$

As seen in Table 3, $R^P M^\bullet$ reaction, which determines the formation of the monomer-radical $R^P M^\bullet$, absolutely dominates, so that only reactions $R^P M_n^\bullet$ and $FR^P M$ do really compete. Since

$$k_{rm}^F > k_p, \tag{7}$$

the monomer conversion graph must include clearly seen IP, similar to that shown for *TEMPO* in Figure 4b. The period is lasting until all the $C_{60}$ content is resumed, thus providing the conversion graph to take the form of a $\Gamma$ letter instead of a rounded hockey stick. Graphs presented in Figure 4c fully confirm these expectations.

Unfortunately, experimental data for the ternary system are not available, but the data in Table 3 related to the addition of *TEMPO* suggest that $SR^P M$ should be the main reaction provided with this radical. As a result, its influence on the FRCP of styrene with both C$_{60}$ and *TEMPO* will either not be noticed at all or lead to an extension in time of the IP, presented in Figure 4c, depending on how competitive is $SR^P M$ reaction with respect to $FR^P M$.

Assessing the results of the above comparative analysis, we can conclude with deep confidence that the virtual representation of the initial phase of free radical polymerization of styrene and its copolymerization with the addition of stable radical *TEMPO* and C$_{60}$ in terms of virtual PPs has stood the test of empirical reality. More detailed concluding comments are given in Section 6.

## 4. Methyl methacrylate polymerization

### 4.1. Digital polymerization passport of methyl methacrylate

According to empirical reality to be discussed below, reaction solution for the FRP and/or FRCP of methyl methacrylate is similar to that of styrene and contains monomer itself, initiator $AIBN$, and small additives of *TEMPO* and C$_{60}$ fullerene [36,37]. Its two-page virtual PP is represented with Table 4 and Figure 5. Nomination of both elementary reaction and DTs of their final products follow listed in Table 1. As previously, yellow marking detects data related to the monomer FRP, while light blue and faint-pink does the same for the monomer FRCP with *TEMPO* and C$_{60}$ fullerene, respectively. Figure 5 presents a standard set of DT images listed in Table 4. Nominations and all the cells content are discussed in details in Section 3.1. The photo-page of the passport is completed with TD images, defined in Table 1 and having the same meaning for all VRSs that include vinyl monomers, free radicals and stable radicals *TEMPO* and C$_{60}$ fullerene. At the same time, the composition of DT samples for different VRSs may differ, thereby reflecting different degree of study of each of the polarizable systems.

The PP photo-page presented in Figure 5 is opened with Headings set (a) related to Table 4, which in contrast to the styrene case involves two DTs associated with $R^A M^\bullet$ just to draw the reader's attention to the need for such verification of the radical activity of the carbon atoms of the vinyl group of the monomer in each case. As seen in Table 4, $E_{cpl}$ of the two DTs differs drastically, thus pointing to the left structure to be considered as acting monomer-radical of the considered VRS. This TD alongside with the DTs set in Figure 5b presents a regular growth of the polymer chain $R^A M_n^\bullet$. All other participants of the considered reaction solution are exhibited in Figures 5c and 5d.

### 4.2. Virtual and real kinetics of the methyl methacrylate polymerization

Thermodynamic descriptors listed in Table 4 show that the combination reactions, laying the foundation of final products $R^A M^\bullet$, $R^A M_n^\bullet$. $SR^A M$, $FM$, $FR^A M$, and $FR^A$ are energetically favorable when the remaining $SR^A$, $FM^\bullet$ and $FS$ are not. Additionally, according to standard energy graphs, they are characterized with inequalities of either $E_{ad} > E_{ac}$ or $E_{ad} \gg E_{ac}$ type, because of which these reactions will not be considered later.

**Table 4.** Elementary reactions and DTs of their final products supplemented with virtual thermodynamic and kinetic descriptors related to the FRCP of methyl methacrylate with stable radicals *TEMPO* and $C_{60}$

| | $R^A M^{\bullet 1)}$ | AIBN$^\bullet$ $R^{A \bullet 1)}$ | TEMPO $R^{T \bullet}$ | $C_{60}$ $F$ | |
|---|---|---|---|---|---|
| | | | | 2-dentate | 1-dentate |
| $M$ | $R^A M_2^{\bullet \bullet}$ | $R^A M^\bullet$ | $R^T M^\bullet$ | $FM$ | $FM^\bullet$ |
| $R^A M^\bullet$ | | | $R^T R^A M$ | $FR^A M$ | |
| $R^{A \bullet}$ | - | - | $R^T R^A$ | $FR^A$ | |
| $R^{T \bullet}$ | - | - | - | $FR^T$ | |
| **Thermodynamic descriptors $E_{cpl}$, kcal/mol** | | | | | |
| $M$ | -15.62 (2) $^{1)}$ -6.95 (3) -18.98 (4) -11.55 (5) | -14.56 (C) $^{2)}$ 1.68 (CH$_2$) $N_{DA}$= 0.96 e | not formed | -19.10 | 29.75 $N_{DA}$= 0.96 e |
| $R^A M^\bullet$ | | | --2.79 | -16.11 | |
| $R^{A \bullet}$ | - | - | 5.001 | -20.14 | |
| $R^{T \bullet}$ | - | - | - | 0.98 | |
| **Kinetic descriptors $E_a$, kcal/mol** | | | | | |
| $M$ | 10.46 (2) | 12.28 (1) | not formed | - | $E_{ad} \gg E_{ac}$ |
| $R^A M^\bullet$ | - | - | 17.63 | 11.58 | |
| $R^{A \bullet}$ | - | - | $E_{ad} > E_{ac}$ | 9.40 | |
| $R^{T \bullet}$ | - | - | - | not formed | |

$^{1)}$ Digits in brackets mark the number of monomers in the oligomer chain.
$^{2)}$ Atomic compositions in brackets mark the target location on the vinyl bond of the monomer-radical.

    The kinetic descriptors in Table 4 are determined from the decomposition barrier profiles related to the relevant DTs that are shown in Figure S3. As seen in the table, four reactions may occur in the VRS of the considered content. Two first $R^A M^\bullet$ and $R^A M_n^\bullet$ provide the FRP of the monomer in the absence of stable radicals. Reaction $SR^A M$ is the only one which reflects the presence of *TEMPO*, while $FR^A M$, and $FR^A$ do the same in the presence of $C_{60}$ fullerene. Evidently, reaction $FR^A$ will compete among the last two. Therefore, only three reactions $R^A M_2^\bullet$, $SR^A M$ and $FR^A$ decide the fate of the first steps of the polymerization of methyl methacrylate in the presence of each of the stable radicals or both together. Only the first two of them concern the polymerization directly while the third one affects only the process' rate because of trapping and thus decreasing the number of free radicals. As for $SR^A M$, the reaction terminates the polymerization completely when its rate exceeds that of $R^A M_2^\bullet$ ($k_p$).

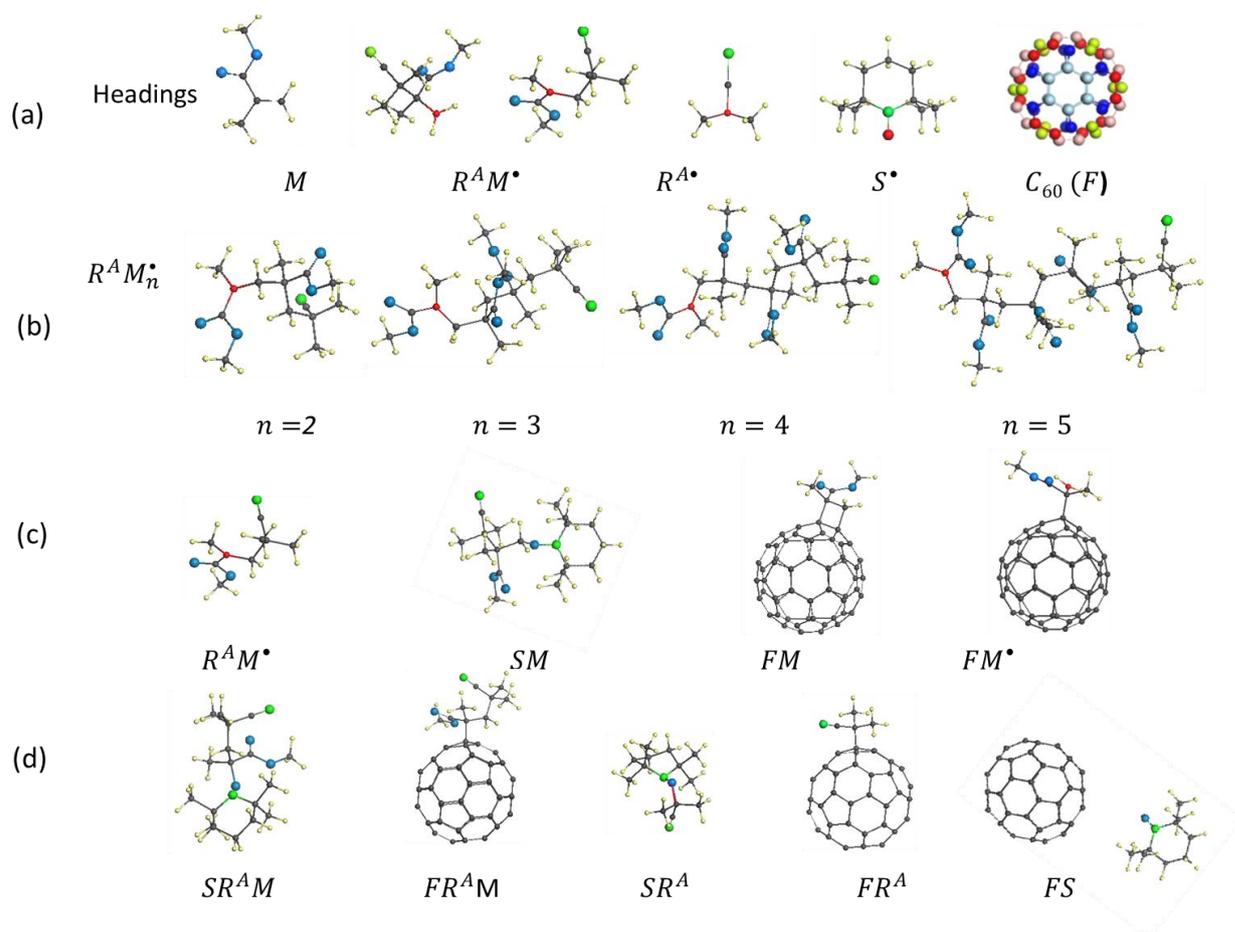

**Figure 5.** Equilibrium structures of digital twins related to the FRCP of methyl methacrylate with stable radicals. (a) Structures, representing headings of Table 4. (b) Oligomer radicals $R^A M_n^\bullet$ for $n$ from 2 till 5 [9]. (c) Digital twins produced by intermolecular interaction of methyl methacrylate with free radical $R^{A\bullet}$, *TEMPO* and C$_{60}$ of both two-dentant ($FM$) and one-dentant ($FM^\bullet$) coupling. (d) DTs, describing the interaction of monomer-radical $R^A M^\bullet$ with *TEMPO* and fullerene C$_{60}$, respectively, as well as DTs related to the interaction of radicals between themselves - $R^{A\bullet}$ with *TEMPO* and fullerene and fullerene with *TEMPO*. Small yellow and gray balls mark hydrogen and carbon atoms, respectively. Larger green and blue balls depict nitrogen and oxygen atoms. Red balls mark carbon (small) and oxygen (large) target atoms. UHF AM1 calculations.

As discussed in the previous section, not kinetic descriptors $E_a$ themselves, but their combinations with the relevant pre-exponential factors $A$ determine the reaction rate. So far, only a qualitative conclusion can be made concerning the latter when looking at atomic compositions of DTs presented in Figure 5. A somewhat more complex composition of $SR^A M$ with respect to that of $R^A M_2^\bullet$ is quite evident, which makes it possible to suggest greater $A$ value for the case thus supposing

$$k_{rm}^S > k_p. \qquad (8)$$

The final conclusion will be made when comparing with experiment. Taking this suggestion in mind, the following prediction of the first steps of the FRP of methyl methacrylate in the presence of stable radicals *TEMPO* and C$_{60}$ fullerene can be made.

The discussed polymerization process, governed with $R^A M^\bullet$ and $R^A M_n^\bullet$ reactions of $k_i$ and $k_p$ rates, respectively, in the presence of *TEMPO* will be terminated and only reaction $SR^A M$ will proceed. It will last until a full content of *TEMPO* is consumed. This should be exhibited in practice by appearing IP in the monomer conversion graph $x(t)$ at the beginning of the reaction, duration of which is progressed when the *TEMPO* content increases. In contrast, the presence of $C_{60}$ fullerene does not touch the polymerization as such and reactions $R^A M_n^\bullet$ and $FR^A$ will occur simultaneously. However, $C_{60}$ affects the rate of the former because of trapping free radicals. The action will be revealed through decreasing slope of the monomer conversion graph $x(t)$. The slope will progress with increasing the fullerene content. In the case of a simultaneous presence of both radicals, the two actions will be superimposed.

The available experimental data allow checking the suggested conclusions. Presented in Figure 6 is a set of monomer-conversion graphs $x(t)$ related to FRP of methyl methacrylate and its FRCP with *TEMPO* and $C_{60}$ fullerene [36,37]. The graphs reveal all three types of behavior mentioned above. Thus, graphs 1-3 in Figure 6a present the monomer polymerization in the presence of $C_{60}$ fullerene. As seen, all the graphs exhibit the straight-line $x(t)$ of the monomer, the rate of which decreases when the fullerene content grows. Addition of *TEMPO* in the reaction solution is followed with the appearance of the IP of the ~1 hour duration (graphs 1 and 1' in Figure 6b). The consequent addition of $C_{60}$ is expectedly followed with decreasing the slope of conversion graphs (graphs 2 and 2' in Figure 6b). Therefore, the comparison with experiments proves the validity of the suggested conclusions and confirms the logic of inequality (8) for the considered VRS.

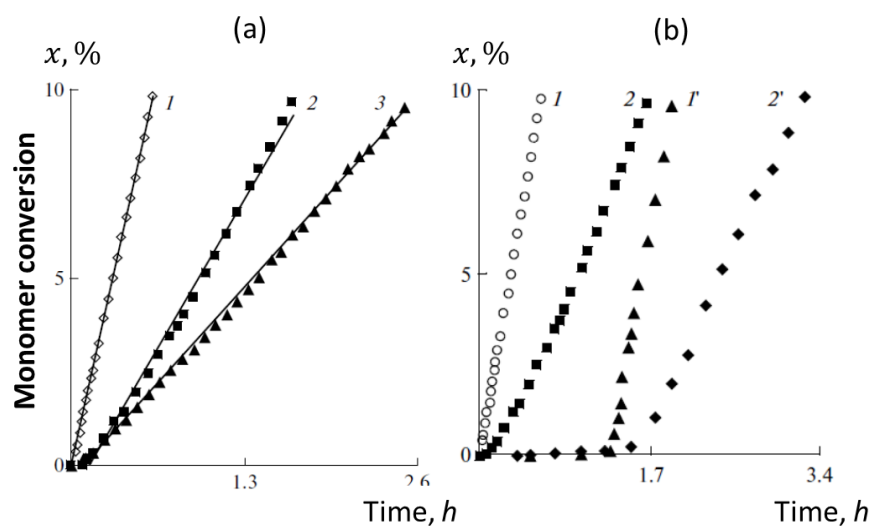

**Figure 6.** Empirical kinetics of the initial stage of both FRP of methyl methacrylate and its FRCP with *TEMPO* and fullerene $C_{60}$. (a) $AIBN^\bullet$-initiated conversion of the monomer in the presence of different $[C_{60}]$: 0 (graph 1); $1.0 \times 10^{-3}$ (graph 2); $2.0 \times 10^{-3}$ mol/L (graph 3). (b) The same but with [*TEMPO* ($C_{60}$)] 0 (graph 1); [*TEMPO*] $1.0 \times 10^{-3}$ mol/L and $[C_{60}]$ 0 (graph 1'); [*TEMPO*] 0 and $[C_{60}]$ $1.0 \times 10^{-3}$ mol/L (graph 2); [*TEMPO* ($C_{60}$)] $1.0 \times 10^{-3}$ mol/L (graph 2'). (T=60$^0$C; o-DCB solvent; [MMA]=2.0 mol/L; [$AIBN$]=2.0x10$^{-2}$mol/L. Digitalized data of Refs. 36 and 37.

## 5. NIPA (*N*-isopropyl acrylamide) polymerization

### 5.1. Digital polymerization passport of NIPA

In this Section we will talk about the NIPA FRP initiated with free radical $AIBN^\bullet$ and its FRCP with $C_{60}$ fullerene. The data listed in Table 5 present the text-page of the monomer virtual PP. The names of the elementary reactions and/or the corresponding DTs are disclosed in Table 1. The photo-page of the passport is given by Figure 7, which presents graphic images of the DTs equilibrium structures, the data for which are given in Table 5. As previously, this table consists of three parts. The first parts contains the nominations, related to both elementary reactions and the relevant DTs that are important for the consideration of the FRP and FRCP under study. The second part involves thermodynamic descriptors $E_{cpl}$ of their final products. The third part presents kinetic descriptors $E_a$ Bold-marked $E_a$ values listed in the table were evaluated when constructing the relevant barrier profiles of the DTs decomposition (see Figure S4).

**Table 5.** Elementary reactions and DTs of their final products supplemented with virtual thermodynamic and kinetic descriptors related to the FRCP of NIPA with fullerene $C_{60}$

| | $R^AM^\bullet$ | $AIBN^\bullet$ $R^{A\bullet}$ | $C_{60}$ $F$ | |
|---|---|---|---|---|
| **Digital twins** | | | | |
| $M$ | $R^AM_2^{\bullet\bullet}$ | $R^AM^\bullet$ | 2-dentate $FM$ | 1-dentate $FM^\bullet$ |
| $R^AM^\bullet$ | $(R^AM)_2$ | $R^AR^AM$ | $FR^AM$ | |
| $R^{A\bullet}$ | - | - | $FR^A$ | |
| **Thermodynamic descriptors $E_{cpl}$, kcal/mol** [1)] | | | | |
| $M$ | -25,378(2) -23.51 (3) -24.05 (4) | −5.675 (1) | -26.483 | 0.598 |
| $R^AM^\bullet$ | - | - | -40.161 | |
| $R^{A\bullet}$ | - | - | -20.137 | |
| **Kinetic descriptors $E_a$, kcal/mol** [1,2)] | | | | |
| $M$ | **8.39 (2)** 7.74 (3) [3)] 8.49 (4) [3)] | **19.09 (1)** | 17.29 (1) [4)] 27.79 (2) | 20.01 |
| $R^AM^\bullet$ | - | - | **0.023** | |
| $R^{A\bullet}$ | - | - | **9.398** | |

[1)] Digits in brackets mark the number of monomers in the oligomer chain.
[2)] Bold data are determined from the barrier profiles presented in Figure S4.
[3)] Data are calculated by using Evans-Polany-Semenov relation presented in [8].
[4)] Digits in brackets indicate one-dentant and two-dentant coupling, respectively.

Main inhabitants of the considered VRS, which are the headings of Table 5 completed with $M$, are given in Figure 7a. NIPA oligomer-radicals, concerning the FRP of the latter (data in the yellow cells of the table) discussed earlier [8] and are not presented in the figure. In contrast, all four members of the NIPA FRCP with $C_{60}$ (the content of faint-pink cells of the table) are given in Figure 7b. The barrier profiles of the decomposition of the latters are shown in Figure S4. Among the latters, the attention should be particularly given to Figure S4a. It exhibits results of the decomposition of fullerenyl $FM$, which is formed by a two-dentant coupling of NIPA molecule with fullerene. The corresponding intermolecular junction is of 2,2-cyclo addition form. Evidently, the junction breaking concerns two $sp^3$C-C bonds. This can happen by elongation and breaking of both bonds either simultaneously, or one-by-one sequentially. These two ways are presented in the figure revealing a quite significant difference in the $E_a$ values in the two cases. Figure S4b presents the barrier profiles of the decomposition of three fullerenyls $FM^\bullet$, $FR^AM$, and $FR^A$, which alongside with the above considered $FM$ constitute a fullerene $C_{60}$ family of DTs, once of interest from the standpoint of the reaction under consideration. Noteworthy is the fact that the $FR^AM$ decomposition turns out to be barrier-free.

### 5.2. Virtual and real kinetics of the FRCP of NIPA with fullerene $C_{60}$

NIPA stands somewhat apart from the bulk of vinyl monomers, which is due to the water solubility of its copolymers with $C_{60}$ fullerene, in contrast to the dominant majority of vinyl polymers, which are insoluble in water. Water-solubility is of particular interest due to the promise of NIPA polymers use in biology and medicine, for example, as nanocontainers for the delivery of drugs to "sick" cells. They can also be used for water purification from microimpurities or extraction from aqueous solutions of valuable substances present in low concentrations. The copolymerization reaction of NIPA with fullerene $C_{60}$ is described in detail elsewhere [34,35]. In this study, special attention is paid to the kinetics of the initial stage of the reaction, represented in terms of the time dependence of monomer conversion $x(t)$ exhibited in Figure 8. As seen in the figure, the evident feature of the dependence concerns the long IP of practically zero amplitude with respect to the abscissa. The IP length is the largest of known for vinyl polymers up today. It means that input of small additive of $C_{60}$ into the chemicl reactor terminates the NIPA polymerization for about 7 hours, after which a rapid occurrence of the process takes place similarly to that observed for styrene or methyl methacrilate.

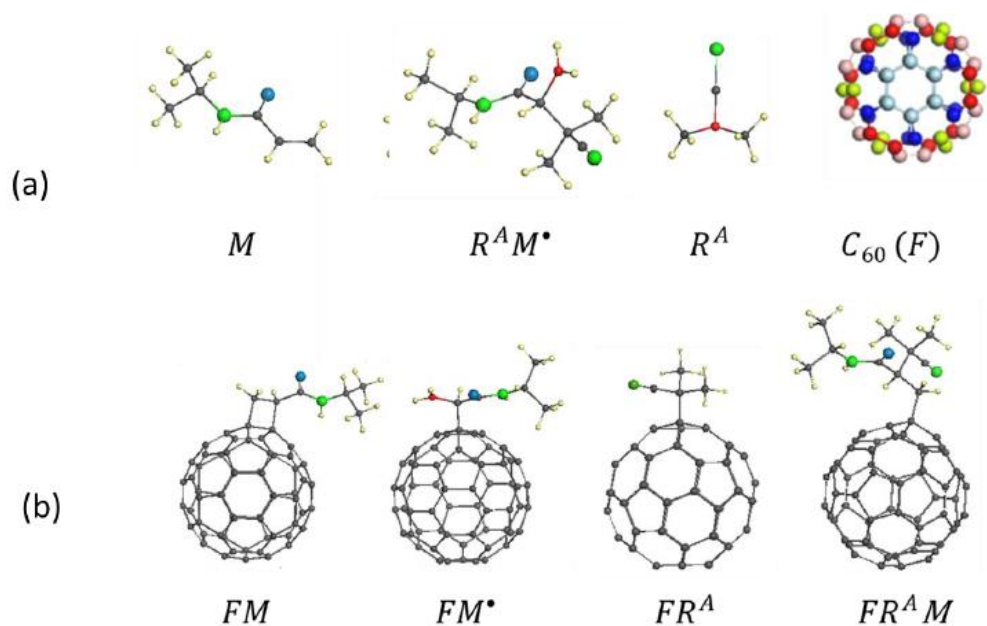

**Figure 7.** Equilibrium structures of digital twins related to the FRCP of NIPA with fullerene $C_{60}$. (a) Digital twins, representing headings of Table 5. (b) Digital-twin fullerenyls produced by intermolecular interaction of $C_{60}$ with NIPA, both two-dentant $FM$ and one-dentant $FM^\bullet$ coupling as well as by trapping of either free radical ($AIBN^\bullet$) $FR^A$ or monomer-radical $FR^AM$. Small yellow and gray balls mark hydrogen and carbon atoms, respectively. Larger green and blue balls depict nitrogen and oxygen atoms. Red balls mark carbon (small) and oxygen (large) target atoms. UHF AM1 calculations.

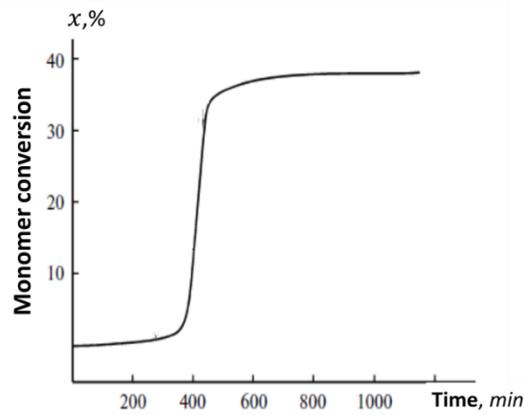

**Figure 8.** Empirical kinetics of the $AIBN^\bullet$-initiated FRCP of NIPA with fullerene C$_{60}$. T=60°C; o-DCB solvent; [NIPA]=0.73 mol/L; [AIBN]=0.24 mol/L; [C$_{60}$]=6.7x10$^{-3}$mol/L. Digitalized data of Ref. 35.

Considering perspectives of the FRCP kinetics presented by descriptors $E_a$ in Table 5, one can conclude that it is this type of reaction, similar to that observed experimentally, that could be expected for NIPA in the presence of small additions of fullerene. Indeed, the active participation of fullerene in copolymerization is determined by three elementary reactions: $FM^\bullet$, $FR^A$, and $FR^AM$. The first reaction determines the polymerization of NIPA on fullerene, caused by the formation of a monomer-radical $FM^\bullet$. As can be seen from the table, the formation of this fullerenyl is accompanied by a positive coupling energy, as a result of which $E_{ad} > E_{aa}$, so that the formation of this fullerenyl from fullerene and NIPA is kinetically unfavorable. In contrast to this case, the formation of fullerenyls $FR^A$, and $FR^AM$ is kinetically welcome, but the huge difference between their descriptors makes it possible to simply forget about the first of them. Thus, in the VRS containing fullerene C$_{60}$, the $FR^AM$ reaction has an evident kinetic advantage, which determines the capture of the $R^AM^\bullet$ monomer-radical by the fullerene and thus stops the polymerization of the monomer. Extremely low value of its descriptor makes this capture practically barrier-free. Accordingly, the following chain of the rate constants related to the fullerene-associated elementary reactions looks like

$$k_{rm}^F \gg k_R^F \gg k_{1m}^F, \tag{9}$$

while the main inequality that controls the beginning of the polymerization in the considered VRS takes the form

$$k_{rm}^F \gg k_p. \tag{10}$$

This is in perfect agreement with the reality presented in Figure 8: the addition of fullerene stops the FRP of NIPA, which resumes after the added C$_{60}$ is fully consumed.

## 6. Conclusive comments

This paper presents an overview of a series of virtual experiments concerning FRP of vinyl monomers and their copolymerization with small additives of stable radicals, such as *TEMPO* and C$_{60}$ fullerene [8-11]. The concept of digital twins, which has successfully demonstrated itself in the virtualization of FRP of vinyl monomers [8], is extended [9-11] to identify reasons of the influence of minor additive of extra stable radicals on the basic FRP process. The success of the implementation of the concept is provided with the reliability of the main idea, which considers polymerization as a chain reaction consisting of many independent elementary reactions occurring in superposition. From this viewpoint, polymerization is the product of won kinetic competition in the environment of a set of elementary reactions. Final products of these reactions alongside with initial ingredients compose VRSs and present a pool of digitalized DTs. The quantum chemical approximation used assumes that coupling energy $E_{cpl}$ of each DT as well as activation energy $E_{ac}$ of the corresponding elementary reaction are thermodynamic and kinetic descriptors that best suit the chemical event digitalization. A matrix type of tabulation is suggested to present the results obtained. The tables, supplemented with the DTs graphical equilibrium structures, represent digital polymerization passport of a VRS, the kinetics of which is under consideration.

In this study, the family of vinyl monomers is presented with styrene, methyl methacrylate and *N*- propyl acrylamide which undergoes rapid polymerization when initiated by the free radicals of either alkyl-nitrile $AIBN^\bullet$ or benzoyl-peroxide $BP^\bullet$. TEMPO ($S$) and C$_{60}$ fullerene ($F$) present stable radicals. Both stimulate the copolymerization of monomers, which has a significant effect on the polymerization kinetics. The FRP of the monomer and its FRCP with stable radicals are separated into elementary-reaction network, each of which was analyzed numerically using a standard energy graph of the reaction theory. The obtained results filled all the fields of digital VRS passports mentioned.

Based on thermodynamic and kinetic descriptors, the digitalization of FRP of vinyl monomers and their FRCP with stable radicals can be considered separately. The former is kinetically controlled with rate constants related to the formation of monomer-radicals $RM^\bullet$ ($k_i$) and the reaction $RM_n^\bullet$ propagation ($k_p$). Both thermodynamic and kinetic descriptors of these reactions obtained in the studies have much in common and exhibit a high efficiency of $AIBN^\bullet$ and $BP^\bullet$ to initiate FRP of all the studied monomers.

In contrast, the introduction of stable radicals into VRSs significantly changes the composition of the elementary reactions involved in polymerization. The latter involve the reactions of capture of monomer molecules $FM$ and $SM$; monomer-radicals $FRM$ and $SRM$ and free radicals by stable ones $FR$ and $SR$, which determines the unfolding kinetic battle in the initial stage of polymerization. As it turned out, the mono-targeted radical *TEMPO* and the multi-targeted C$_{60}$ behave differently in this battle. Kinetic descriptors associated with *TEMPO* highlight the monomer-radical capture reaction $SRM$ as the main one in all the studied cases. Capture of the monomer-radical prevents the beginning or stops the onset of the monomer FRP, which appears on the experimental graph of the time-dependent percentage conversion of the monomer $x(t)$ as a region with zero conversion, known as the induction period (IP). This conclusion of the virtual experiment is fully consistent with the available experimental data concerning styrene and methyl methacrylate (see Refs. 36 and 37 and references therein).

At the same time, analysis of kinetic descriptors shows that the range of elementary reactions imposed by fullerene and subject to accounting is wider and includes several possibilities, the main of which are $FM$, $FRM$ and $FR$. The assumption of independence and superposition of all elementary reactions involved in the polymerization allows for predictive analysis of kinetic winners by comparing their descriptors for similar reactions. In the case of FRCP of a monomer with fullerene, these two groups, with a certain degree of assumption, are the main reactions of FRP $RM^\bullet$ and $RM_n^\bullet$, on the one hand, and the FRCP reactions $FM$, $FRM$ and $FR$, on the other. If in the first case, as already mentioned, the FRP reactions do not differ qualitatively and do not reveal exceptional features caused by change of either monomer or initiating free radical, then small additions of fullerene changes the nature of the polymerization source drastically. As occurred, kinetic descriptors of the above three reactions are highly sensitive to the monomer and initiating free radical chemical composition. Thus, in the series of monomers styrene - methyl methacrylate - NIPA, upon $AIBN^\bullet$ initiation, the fastest reactions are $FM$, $FR$ and $FRM$, respectively. In the case of styrene, we are talking about the $FM^\bullet$ reaction, which corresponds to the one-dentant addition of styrene to the fullerene molecule. The fate of the onset of polymerization is determined by the ratio of the rate constants of the FRP and FRCP reactions. So in the case of styrene, $k^F_{1m} > k_i$, which means that the formation of a fullerenyl monomer radical $FM^\bullet$ is the winner in the competition of intermolecular interactions. The ratio of the rates constants $k^F_{1m}$ and $k^F_p$ is similar to that of $k_i$ and $k_p$ of the basic FRP reaction, so that the radical polymerization of styrene, initiated with fullerene, becomes the first reaction of the polymerization process in the VRS consisting of styrene, *AIBN* and C$_{60}$ fullerene. Since in reality the content of styrene and fullerene in the reaction solution differs by three orders of magnitude, the styrene conversion graph $x(t)$ in the course of its fullerene-based polymerization turns out to be pressed to the time axis on the background of the total conversion, taking the form similar to a routine IP. This circumstance has not yet made it possible to facsimile the polymerization of styrene initiated by fullerene previously, erroneously attributing its $x(t)$ graph to the IP one. The new interpretation of the initial stage of the styrene polymerization in the presence of C$_{60}$ fullerene proposed in this article is confirmed by both virtual and real consideration of the FRP of styrene in the simultaneous presence of two stable radicals *TEMPO* and C$_{60}$.

In the case of methyl methacrylate, capturing free radical $AIBN^\bullet$ with C$_{60}$ is the main reaction promoted with fullerene. The reaction does not affect the polymerization process as such, but provides a reduction of free radicals content thus changing the slope of the $x(t)$ graph pressing it towards the time axes. It is this type of the empirical $x(t)$ graph and its dependence on the C$_{60}$ content, which is characteristic for the empirical monomer conversion in the case of the FRCP of methyl methacrylate with C$_{60}$ fullerene [36].

Barrier-free $FRM$ reaction is a particular feature of the $AIBN^\bullet$-initiated virtual FRCP of NIPA with C$_{60}$ fullerene. Having no other competitors, this reaction dominates in the VRS containing NIPA, free radicals $AIBN^\bullet$ and C$_{60}$ fullerene, not allowing the FRP of NIPA to take place until all the fullerene content is exhausted. Indeed, FRCP NIPA with fullerene is accompanied by the longest time-consuming IP in the empirical $x(t)$ conversion graph. The IP appearance in the $x(t)$ conversion graph can be stimulated with changing not only monomers, but initiating free radical as well. Thus, when the content of the styrene VRS is changed by substituting *AIBN* with benzoyl peroxide, $FM^\bullet$ fades into the background, and reaction $FRM$ takes first place. In this case, $k^F_{rm} > k_P$, so that the addition of fullerene to the reaction solution containing styrene and benzoyl peroxide terminates the FRP of styrene until the fullerene content is exhausted. It is the situation, which is observed empirically [36].

The discussion presented is not exhaustive and leaves many topics for further detailed consideration. Thus, all the considered reactions stimulated by fullerene are limited to the formation of monofullerenyls. In fact, fullerene is very prone to polyderivatization [56], according to which each of the fullerene reactions listed in Table 1 should be considered for potential production of polyfullerenyls. The number of such substances is huge, so completing the digitalization of a polymer product will require the creation of an algorithm for the kinetic selection of polyderivatives of the fullerene molecule similar to a thermodynamic algorithm using the features of the spin density of fullerene [55,56]. The search for such an algorithm will require significant joint efforts of experimenters and computer scientists. So far, the author's goal was to draw attention to the fact that the multifaceted picture of free-radical polymerization of vinyl monomers, at least in the initial stage, can be reliably certified using the latest methods of digitalization of this process and freely extended to other polymerized systems. The polymerization-passport approach occurs highly productive thus allowing sharing friendly the responsibility of the polymerizable system study between virtual and real experiments. Modern quantum chemistry of radicals is able to construct required passports within wide limits.

Consequently, digitalization of polymerization reactions may prove to be a broad platform for introducing this approximation into polymer science and technology.

**Acknowledgments.** The author is thankful to E.G. Atovmyan for deep discussions that have stimulated interest to free-radical polymerization of vinyl monomers and have attracted the author's attention to a particular role of $C_{60}$ fullerene in this chemical event. This paper has been supported by the RUDN University Strategic Academic Leadership Program.

# Digitalization of Free-Radical Polymerization

Elena F. Sheka

Institute of Physical Researches and Technology, Peoples' Friendship University of Russia (RUDN University), 117198 Moscow, Russia;

sheka@icp.ac.ru

SUPPORTING INFORMATION

Decomposition barrier profiles were exploited to manifest main energetic parameters of studied elementary reactions. Figures S1 and S2 present selected examples of standard and complicated situations met under the way of the digitalizing of various barrier profiles related to the styrene polymerization. A typical situation, when all the features of the energy graphs $E(R)$ are well defined, is shown in Figure S1. Two main extremes of the plottings attributed to their minima and maxima are clearly vivid, thus providing a direct evaluation of $E_{ad}$ and $E_{ac}$ values. The examples given relate to the first two members of the styrene polymer chain $FM_n^\bullet$ ($FM^\bullet$ and $FM_2^\bullet$) attached to the fullerene and to the $FM$ describing the styrene two-dentant attaching to the fullerene. Noteworthy are the well-defined maxima of the graphs, which record the energy of the transition state $E_{TS}$ of the products of the corresponding elementary reactions. In all cases, the intermolecular junctions in the products, which play the role of reaction coordinate, are determined by an *sp³*C-CH bond, which is broken in the transition state [4]. The latter was evidenced by the fact that the position of the energy graphs maxima coincides with $R_{crit}^{C-C}$ of 2.11±0.1 Å that determines maximum length of the *sp³*C-C bond, above which the bond becomes radicalized thus revealing the start of its breaking [1-4]. Oppositely to the case, the *sp³*C-O bond forms the junctions in $SFM$ and $SR^AM$ products presented in Figure S1b. Expectedly, the maximum positions of their barrier profiles should differ from those provided with breaking the *sp³*C-C bond that is a reaction coordinate of the $FR^AM$ decomposition. Actually, these positions constitute 1.91 and 2.21 Å, which well correlates with $R_{crit}^{C-O}$ of 2.01 Å relating to the dissociation of the *sp³*C-O bond of ethylene glycol presented in the bottom of the figure. Evidently, $R_{crit}^{C-O}$, as well as $R_{crit}^{C-C}$, deviates in different atomic surrounding, which was observed in the current study. Shown in Figure 5c is related to the case, when the determination of $E_{TS}$ becomes uncertain. Two DTs presented in the figure have the same atomic compositions while formed (and reverse decomposed) differently. The first DT $FR^AM$ is the result of capturing monomer-radical $R^AM^\bullet$ with fullerene. Accordingly, the intermolecular junction is formed by the *sp³*C-C bond, marked with red, that is the reaction coordinate when the product decomposes. In contrast, DT $R^AFM$ presents the case when monomer-radical $FM^\bullet$ is inhibited with free radical $R^{A\bullet}$. The intermolecular junction as well as the reaction coordinate is transmitted to the *sp³*C-O bond. The corresponding bonds are red-marked. As seen in the figure, if the determination of the corresponding energy $E_{TS}$ is not difficult in the first case, in the second case it becomes uncertain.

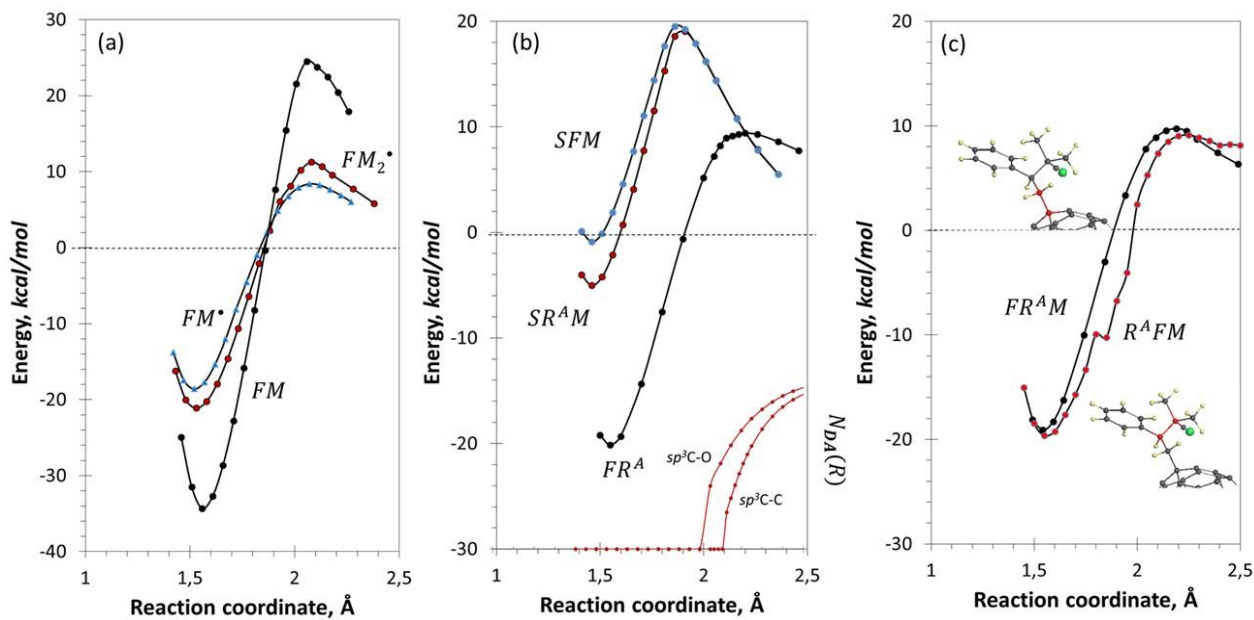

**Figure S1.** Virtual barrier profiles of the decomposition of DTs participating in the FRP of styrene and its FRCP with *TEMPO* and $C_{60}$ fullerene. DTs are to the following final products of the studied elementary reactions. (a) Two first members of the styrene polymerization on $C_{60}$ fullerene $FM_n^\bullet$ ($FM^\bullet$ and $FM_2^\bullet$) as well as fullerenyl formed by two-dentant attachment of styrene ($FM$). (b) Monomer-radicals capturing of $FM^\bullet$ and $R^A M^\bullet$ with *TEMPO* ($SFM$ and $SR^A M$) as well as of $R^A\bullet$ radical with fullerene ($FR^A$). (c) Capturing of $R^A M^\bullet$ monomer-radical with *TEMPO* ($SR^A M$) while of $FM^\bullet$ one with free radical $R^A\bullet$ ($R^A FM$). $N_{DA}(R)$ graphs are related to the dissociation of the $sp^3$C-C bond of ethane and $sp^3$C-O bond of ethylene glycol [1-3]. UHF AM1 calculations.

A particular behavior of the *sp³*C-O and its inability to play the role of the reaction coordinate is justified by the players of the styrene FRP game when the initiating free radical $AIBN^\bullet$ is replaced with $BP^\bullet$. Figure S2 presents two pairs of DTs where this coordinate of one partner (either $R^PM_n^\bullet$ (a) or $FR^PM$ (b)) is played with the *sp³*C-C bond, while the *sp³*C-O bond plays the role for other partners ($R^PM^\bullet$ (a) and $R^PFM$ (b)). As seen in the figure, in the first case the $E_{TS}$ values are well evaluated and localized within the region of the expected values for $R_{crit}^{C-C}$. In the second case, this evaluation becomes uncertain. Evidently, the radical character of the considering elementary reactions is the feature's reason. However, more detailed answering the question requires further investigation.

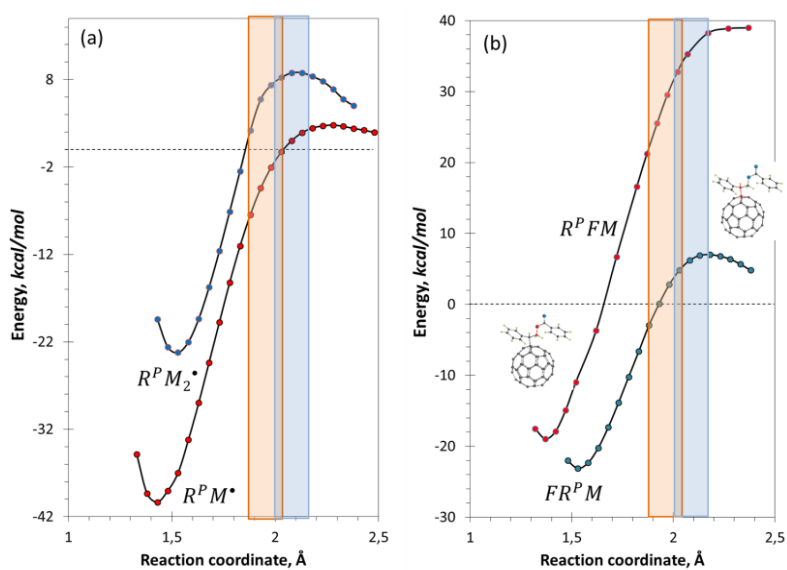

**Figure S2.** The same but initiated with benzoyl peroxide. DTs are to the following final products of the studied elementary reactions. (a) Monomer- and dimer-radicals $R^PM^\bullet$ and $R^PM_2^\bullet$, respectively. (b) Products of capturing $R^PM^\bullet$ monomer-radical with fullerene ($FR^PM$) while of $FM^\bullet$ one with free radical $R^{P\bullet}$ ($R^PFM$). Light pink and light blue bands mark assumed deviation intervals of critic values, $R_{crit}^{C-O}$, and well as $R_{crit}^{C-C}$ of the maximum length of the $sp^3$C-C and $sp^3$C-O bonds under breaking. UHF AM1 calculations.

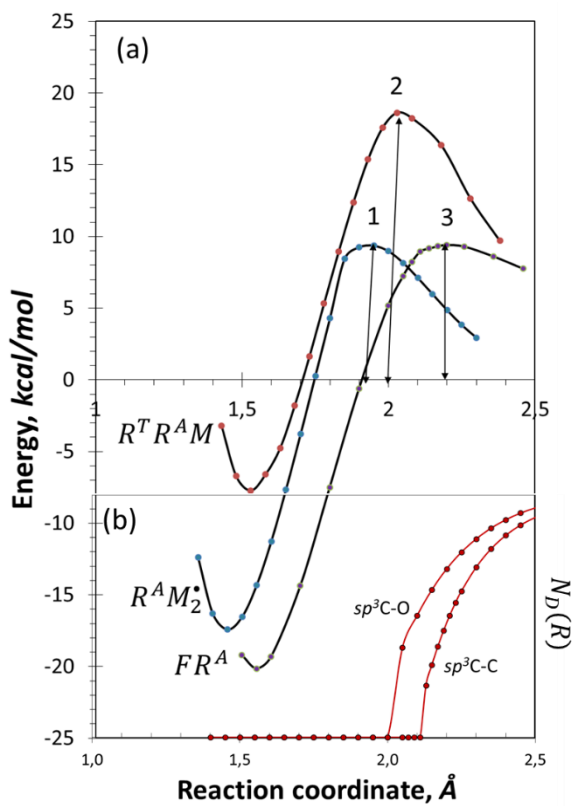

**Figure S3.** The same as in Figure S1, but for methyl methacrylate. (a) Dimer radicals $R^A M_2^{\bullet}$ (1), capturing products $R^T R^A M$ (2) and $F R^A$ (3). (b). $N_D(R)$ graphs related to the dissociation of the $sp^3$C-C bond of ethane and $sp^3$C-O bond of ethylene glycol [1-3]. UHF AM1 calculations.

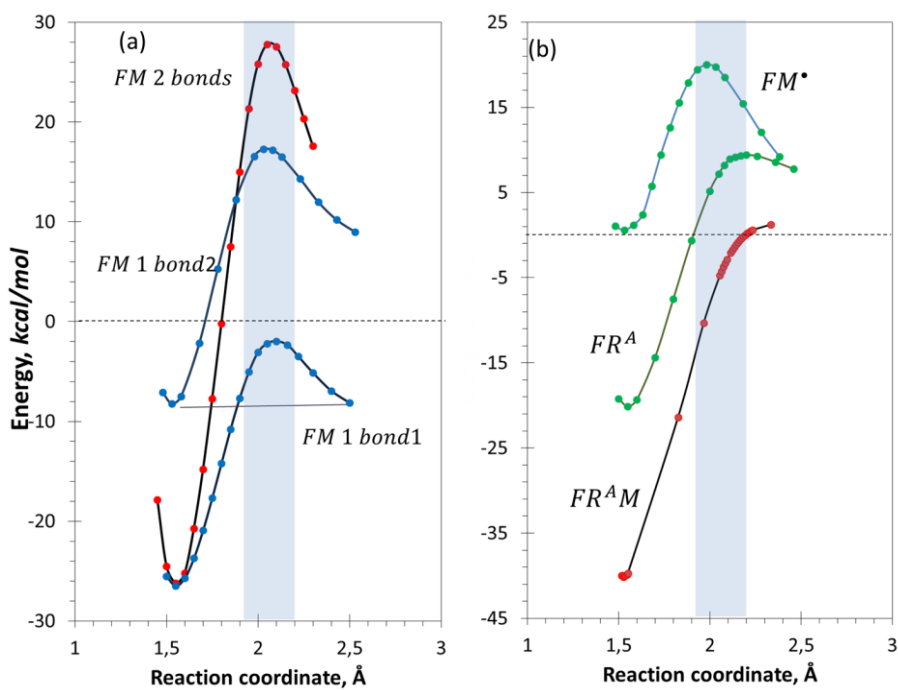

**Figure S4.** The same as in Figure S1, but for *N*-isopropyl acrylamide (NIPA). (a) Fullerenyl $FM$, related to two-dentant coupling of NIPA with fullerene and decomposed through elongation of either two bonds simultaneously, or through one-by-one bond sequentially. (b) Fullerenyls $FM^\bullet, FR^A$ and $FR^AM$ related to the elementary reactions of the same name. UHF AM1 calculations.